\documentclass{pnastwo}

\usepackage{amsmath}
\usepackage{accents}
\usepackage{graphicx}
\usepackage{natbib}
\usepackage{wrapfig}
\usepackage{subfigure}
\usepackage{nameref}

  \newcommand{\fig}[1]{Fig.~\ref{#1}}

\begin{document}

\title{The Exoplanet Orbital Eccentricity -
Multiplicity Relation and the Solar System}

\author{Mary Anne Limbach$^1$\affil{1}{Department of Astrophysical 
Sciences,
Princeton University, Princeton, NJ 08544, USA} \affil{2}{Department of
Mechanical
and Aerospace Engineering, Princeton University, Princeton, NJ 08544, USA}
\and
Edwin L. Turner\affil{1}{}\affil{3}{The Kavli Institute for the Physics 
and
Mathematics of the Universe, The University of Tokyo, Kashiwa 227-8568,
Japan}}

\maketitle
\footnotetext[1]{Published previously under Mary Anne Peters}

\begin{article}
\begin{abstract}
The known population of exoplanets exhibits a much wider range of orbital
eccentricities than Solar System planets and has a much higher average
eccentricity.
These facts have been widely interpreted to indicate that the Solar
System is an
atypical member of the overall population of planetary systems.  We
report here on a
strong anti-correlation of orbital eccentricity with multiplicity (number
of planets
in the system) among catalogued radial velocity (RV) systems.  The mean, 
median and rough
distribution
of eccentricities of Solar System planets fits an extrapolation of this
anti-correlation to the eight planet case rather precisely despite the 
fact that no
more than 2 Solar System planets would be detectable with RV data 
comparable to that
in the exoplanet sample.  Moreover, even if regarded as a single or double 
planetary system, the Solar System lies in a reasonably heavily populated 
region of eccentricity-multiplicity space.
Thus, {\it
the Solar
System is not anomalous among known exoplanetary systems with respect to
eccentricities when its multiplicity is taken into account}.
Specifically, as the
multiplicity of a system increases the eccentricity decreases roughly as a
power law
of index --1.20.  A simple and plausible but {\it ad hoc} and 
model-dependent
interpretation of this
relationship
implies that approximately 80\% of the one planet and 25\% of the two 
planet
systems in our sample have additional, as yet undiscovered, members
but that systems of higher observed multiplicity are largely complete
({\it i.e.}, relatively rarely contain additional undiscovered  planets). 
If low
eccentricities indeed favor high multiplicities, habitability may be more 
common in systems
with a larger number of planets.
  \end{abstract}


%%%%%%%%%%%%%%%%%%%%%%%%%%%%%%%%%%%%%%%%%%%%%%%%%%%%%%%%%%%%%%%%%%%%%%
\section{Significance Statement}\label{sec:Significance}
The Solar System planets have near-circular orbits (i.e. unusually low 
eccentricity)
compared with the known population of exoplanets, planets that orbit stars 
other
than the Sun. This fact has been widely interpreted to indicate that the 
Solar
System is an atypical member of the overall population of planetary 
systems.  We
find a strong anti-correlation of orbital eccentricity with the number of 
planets
(multiplicity) in a system that extrapolates nicely to the eight-planet, 
Solar
System case despite the fact that no more than 2 Solar System planets 
would be
detectable in the sample in which the anti-correlation was discovered. 
Habitability
may be more common in systems with a larger number of planets which have 
lower
typical eccentricities.

%%%%%%%%%%%%%%%%%%%%%%%%%%%%%%%%%%%%%%%%%%%%%%%%%%%%%%%%%%%%%%%%%%%%%%
\section{Introduction}\label{sec:Introduction}

Solar System orbital eccentricities are unusually low compared to those of
exoplanets. This fact is one of the most
frequently
noted major surprises revealed by the discovery and early explorations of 
the
exoplanet population orbiting Sun-like stars and has been widely
interpreted to
indicate that the Solar System is not a representative example of a
planetary system
(reviews by \cite{Marcy2005Observed,Udry2007Statistical, 
Perryman2011Exoplanet} and
references therein).  Many planetary formation theories developed prior to 
the
discovery of exoplanets suggested planets would have eccentricities 
similar to the
Solar System planets \cite{Lissauer1995Urey, Boss1995Proximity}. Several 
attempts
have been made to accurately model the dynamical
evolution
of planetary systems since then with the goal of explaining the observed 
eccentricity
distribution \cite{Rasio1996Dynamical, Weidenschilling1996Gravitational,
Adams2003Migration, juric2008DynamicalApJ, juric2008dynamical}. These
papers invoke
planet-planet interactions as the primary mechanism determining the
distribution of
orbital eccentricities.  The most recent of these papers,
\cite{juric2008dynamical},
concludes that there would be a dependence of eccentricity on multiplicity
(the
number of planets in the system) in this scenario. We use existing RV
exoplanet data
to test that prediction.

Our dataset consist of 403 of the 441 cataloged radial velocity (RV)
exoplanets
obtained since the 1990s ({\it exoplanet.org}).  Of these 127 are members
of known
multiple-planet systems with multiplicities of up to six.  The data are
sufficient
to allow an estimate of the relationship of eccentricity to multiplicity.
It has
been noted that eccentricity in two planet systems tends to be lower than
in single
planet systems \cite{Rodigas2009Which}.  This paper explores the
relation
at higher multiplicities and notes its unexpected and surprising 
consistency with the
Solar System case.

The dataset is discussed in the next section. We then show the trend in
eccentricity
with multiplicity and comment on possible sources of error and bias.  Next 
we
measure the mean, median and probability density distribution of
eccentricities for
various multiplicities and fit them to a simple power-law model for
multiplicities
greater than two.  This fit is used to make a rough estimate of the number
of higher
multiplicity systems likely to be contaminating the one and two planet 
system
samples due to as yet undiscovered members under plausible, but far from 
certain, assumptions.  Finally, we
conclude with some
discussion of the implications of this result.

%%%%%%%%%%%%%%%%%%%%%%%%%%%%%%%%%%%%%%%%%%%%%%%%%%%%%%%%%%%%%%%%%%%%%%
\section{The Dataset}
\label{sec:Dataset}

Only radial velocity (RV) exoplanet data obtained from {\it exoplanet.org}
are used for this analysis. All of the RV exoplanets
listed on the
website that have a measured eccentricity are included in the analysis. If
the
eccentricity of the planet was not listed or if it was given as zero, the
exoplanet
was excluded from our sample. Thirty-eight systems were excluded on the
latter
basis, of which twenty-nine had their eccentricity constrained to zero in 
the
orbital fit.  Table \ref{Table:NumPerBin}1 lists the number of RV 
planets in each multiplicity bin.

The 5- and 6-planet systems, one of each, were combined into one bin so
that there
was sufficient data for a statical analysis. Note that the total number of
planets
with a given multiplicity is not necessarily a multiple of the
multiplicity of the
system because not all planets in some systems have measured
eccentricities.  In the
cases where a certain parameter of an exoplanet or exoplanet system is 
under
debate,
we used the {\it exoplanet.org} value.  For instance, the multiplicity of
a system
may be three planets based on observations, however the motion of those 
three
planets may imply that there are additional companions and therefore the
multiplicity of the system may be listed as four planets rather than three
on {\it
exoplanet.org}. In all such cases, we simply adopt the data listed on the
website
for consistency. The data were taken from the website over the course of
several
weeks in February and March 2014. We used RV data only for our
analysis
since the planets in that subset of the  data set typically have known and 
relatively
reliably
measured eccentricities.

%%%%%%%%%%%%%%%%%%%%%%%%%%%%%%%%%%%%%%%%%%%%%%%%%%%%%%%%%%%%%%%%%%%%%%
\section{A Trend in Multiplicity verses Eccentricity}\label{sec:Trend}

\begin{figure}[h]
\centering
\includegraphics[scale = 0.5]{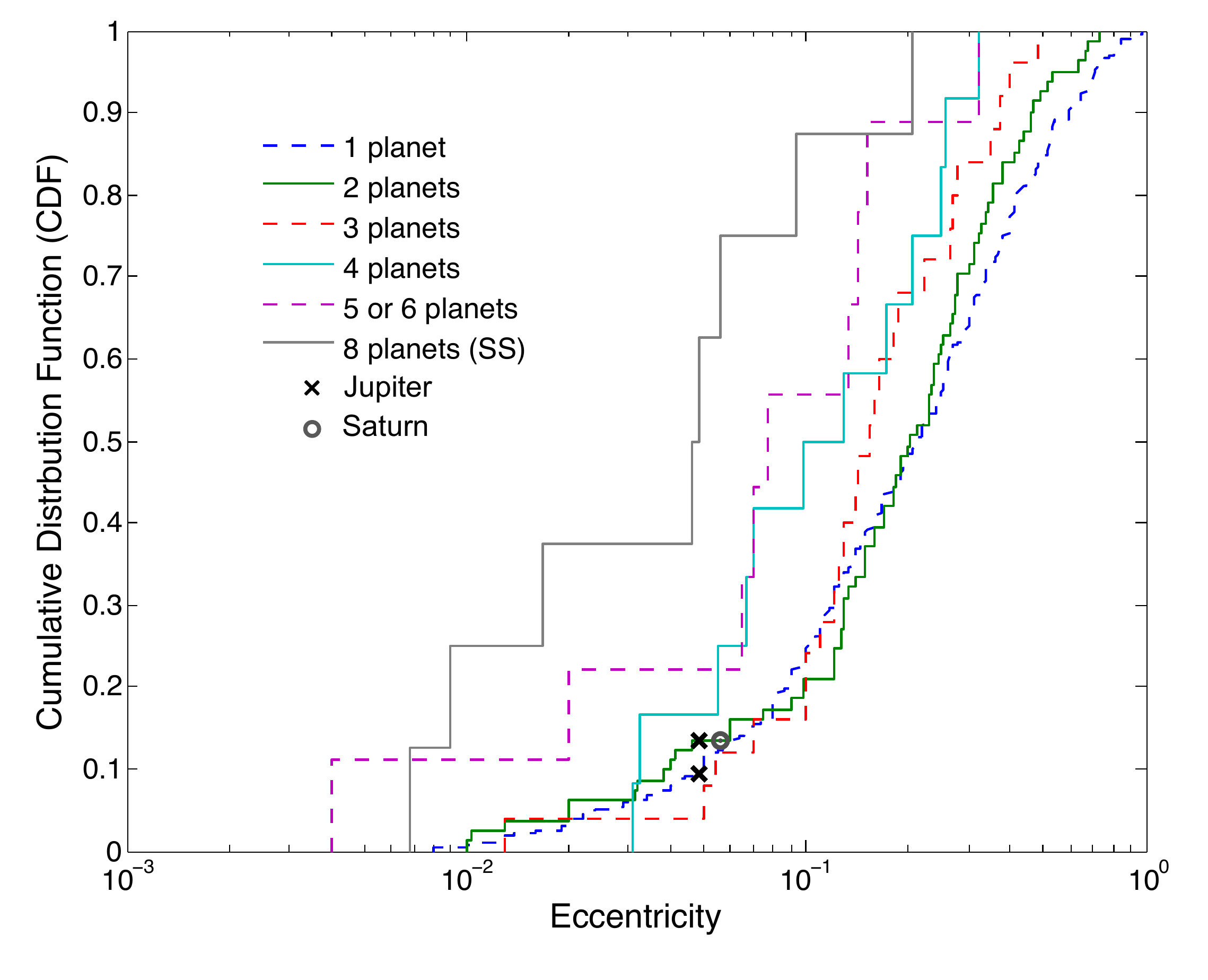}
\caption{Cumulative eccentricity distributions in RV exoplanet systems and 
the Solar
System for
various multiplicities. There is a trend towards lower eccentricities at
higher
multiplicities. The abbreviation `SS' is for `Solar System' planets. A 
black x is
shown for where Jupiter would appear on the 1- and 2-planet distributions, 
and a
gray circle is shown for Saturn on the 2-planet curves. This demonstrates 
that even
if the Solar System was detected via RV as a one or two planet system, it 
would
still be consistent with the data.}
\label{FracObjEcc}
\end{figure}

\begin{figure}[h]
\centering
\includegraphics[scale = 0.46]{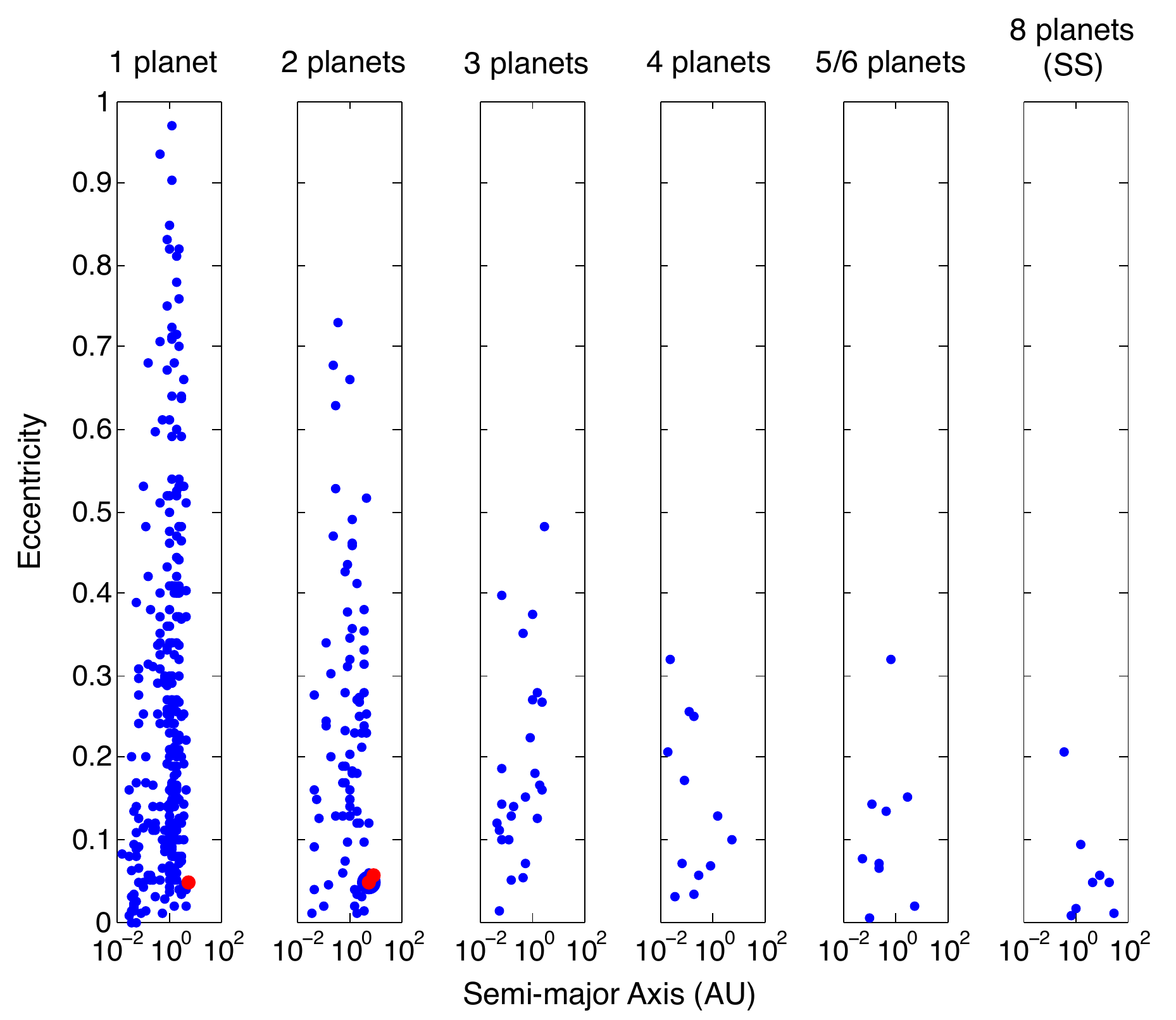}
\caption{Eccentricity verses semi-major axis going from low- (left) to
high-multiplicity (right). The `SS' denotes `Solar System' planets. A red 
dot is
shown for where Jupiter would appear on the 1- and 2-planet distributions, 
and for
Saturn on the 2-planet distribution. This demonstrates that even if the 
Solar System
was detected via RV as a one or two planet system, it would still be
consistent with
the data.}
\label{eccVSau}
\end{figure}

\fig{FracObjEcc} shows the cumulative eccentricity distribution function 
in
exoplanet systems and the Solar System for systems of various
multiplicities. There
is a clear trend towards lower eccentricities in higher multiplicity
systems. This
trend is perhaps most noticeable in the top (high eccentricity) half of
the plot.
\fig{eccVSau} shows the eccentricity vs. semi-major axis relation at each
multiplicity and also displays the strong tendency towards lower
eccentricity at higher multiplicity.

\begin{figure}[h]
\centering
\includegraphics[scale = 0.46]{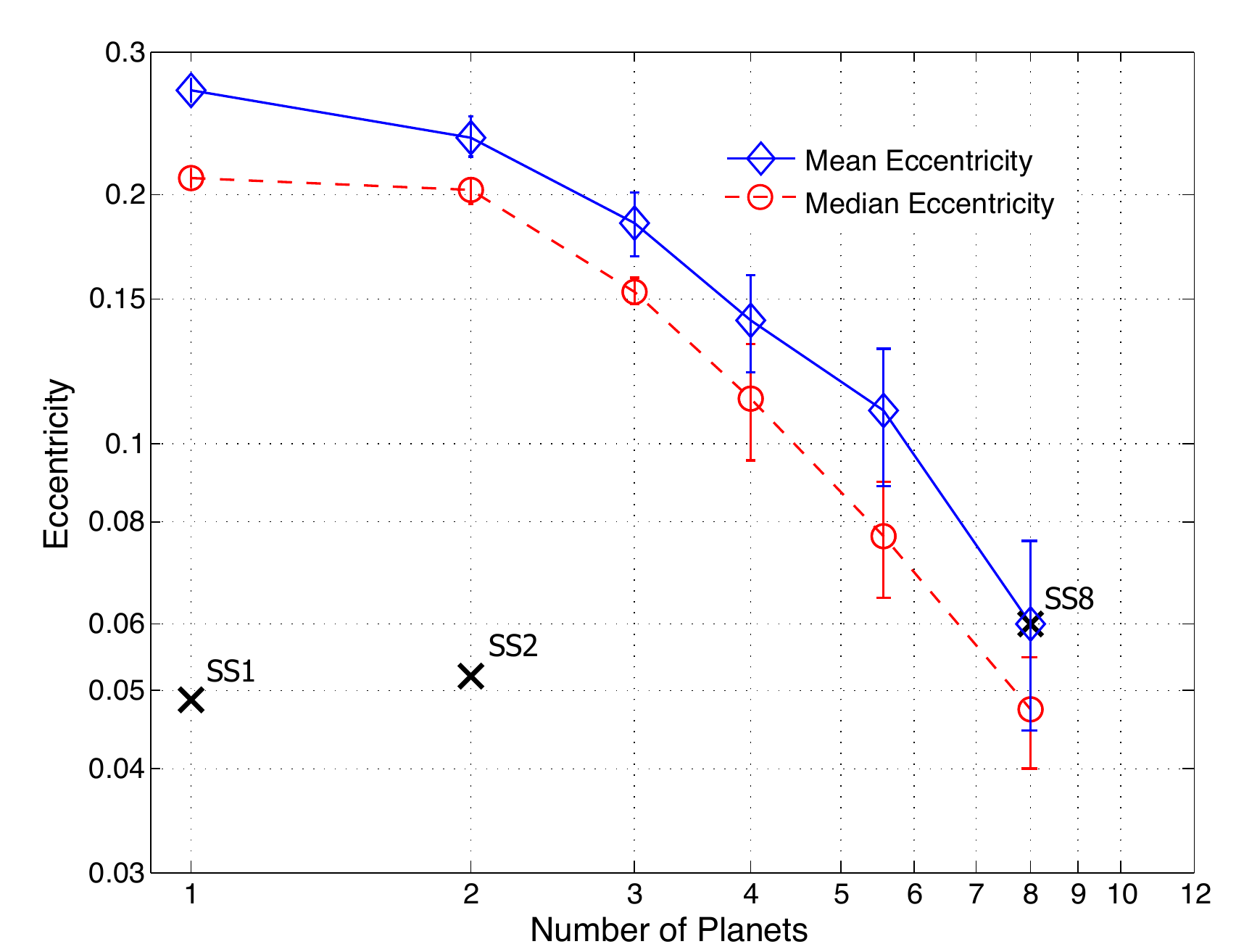}
\caption{Mean and Median Eccentricities in RV exoplanet systems and the
Solar System
as a function of multiplicity (number of planets in the system). As the 
number of
planets increases,
eccentricity
decreases. A plateau in eccentricity at low multiplicity is noticeable
especially in
the median. This is possibly due to contamination of the one-planet data 
with
higher-multiplicity systems. The `SS' denotes `Solar System' planets. If 
the Solar
System was detected via RV, it is mostly likely that only Jupiter (1
planet) or
Jupiter and Saturn (2 planets) would be detected. The mean eccentricity in 
these
scenarios is marked on the plot by black x's and labeled SS1 and SS2, 
respectively.
If the entire Solar System was detected it would appear in the 8-planet 
bin -- this
case is labeled  SS8. Note that the SS8 point is consistent with the the 
trend in
the mean whereas the SS1 and SS2 points are not (which should suggest to a 
hypothetical alien RV exoplanet hunter studying the Solar System from afar 
that it contains more than
one or two planets).}
\label{MeanMedEcc}
\end{figure}

The strength and nature of the anti-correlation of orbital eccentricity 
with
multiplicity is even more dramatically revealed by plotting the mean and
median
eccentricity as a function of multiplicity, as shown in \fig{MeanMedEcc}.
Two features of this figure are worthy of special note:  1 - The
divergence of the one-planet systems, and two-planet systems to a lesser
extent,
from what appears to be a power-law relation at higher multiplicities is
noticeable
in both the mean and median curves.  2 - The Solar System fits the
eccentricity
trend at higher multiplicity quite well despite the fact that the RV data 
quality
for the exoplanetary systems would only allow the detection of Jupiter and 
perhaps
Saturn.  In other words, it would be more statistically consistent to plot 
the SS
point in these figures at a multiplicity of one or two, as also shown in 
the figure.
In either case, the Solar System does not appear
to have
unusually low orbital eccentricities with respect to exoplanet systems,
but rather {\bf the Solar System planet eccentricities are consistent with 
the 
exoplanet
eccentricity distribution when multiplicity is taken into account}.

The uncertainties of the mean eccentricities shown in \fig{MeanMedEcc} 
were
calculated by bootstrapping. The bootstrap method gave an uncertainty of
approximately two-thirds of the usual rms estimator, which is consistent
with the
limited extent of the eccentricity distribution tail. The uncertainties in
the medians
correspond to the one-third and two-thirds points in the distributions
shown in
\fig{FracObjEcc} divided by $\sqrt{N-1}$, where $N$ is the number of
points in the
multiplicity bin.

A 2-sample Kolmogorov--Smirnov (K--S) test was used to determine the
significance of the
differences in the eccentricity distributions of systems with different
multiplicities.
The test was applied both to the entire data sample and to the subsamples
consisting of
the highest 75\% of the eccentricities in each multiplicity subsample.
Use of the latter
subset of the data is motivated by the possibility that low multiplicity
systems with
low eccentricity orbits may have undiscovered members and thus actually be
of higher
multiplicity.

In both cases, the p-values consistently decrease for
larger difference in the multiplicities of the two samples being compared.
This is
consistent with the systematic trends visible in \fig{FracObjEcc} and
\ref{MeanMedEcc}. The results of the tests for the full sample and the 
high
eccentricity
subsamples are shown in Table \ref{Table:KStest}2.
In high eccentricity subsample case, for samples which have multiplicities 
that
differ by at
least two planets,
the K--S test
yield a statistically
significant difference (p $\leq$ 0.05). At higher multiplicities, the
significance
(1 -- p) of the difference between distributions with adjacent
multiplicities
generally decreases. This may well be an artifact of the smaller sample
sizes at higher
multiplicities rather than an actual convergence of the eccentricity
distributions.

\subsection{Sources of Error}\label{Errors}
Systematic errors in the eccentricity distribution of exoplanets are
possibly due to the
techniques used for measuring this parameter and biases associated with
exoplanet
detection methods. Shen \& Turner \cite{Shen2008OntheEccentricity} studied 
the
systematic
bias
of RV determinations of eccentricity using the conventional $\chi^2$
fitting and
showed that such measurements are biased high and significantly so for
low values of $N^{1/2}K/\sigma$, where $K$ is the velocity semi-amplitude
and $\sigma$ is
the typical
uncertainty of each of the $N$ velocity measurements.
One might then be concerned that higher
multiplicity systems receive more telescope time and thus have a higher
value of this parameter and a more accurately determined eccentricity
values
than lower multiplicity systems. If this were the case, one
might observe a multiplicity-eccentricity trend due to observational bias
rather
than a true physical phenomena.
To ensure that this effect is negligible for our dataset,
we conducted a careful analysis of the measured  $N^{1/2}K/\sigma$ values 
and found
that there is no significant effect in our sample.
Furthermore, higher-multiplicity systems
tended to have  $N^{1/2}K/\sigma$ values similar to those of 
lower-multiplicity
systems.
Shen \& Turner \cite{Shen2008OntheEccentricity} only expect eccentricity 
to increase
for
$N^{1/2}K/\sigma \gtrsim 15$ and most of the measurements in our sample 
have higher
$N^{1/2}K/\sigma$
value than
that threshold. For these reasons, we do not
believe that the
multiplicity-eccentricity trend we reported here is influenced by this
potential bias.

A probable shortcoming of the data is undetected companions, especially in
low-multiplicity systems. Given the eccentricity-multiplicity relation,
this would add low
eccentricity planets to the low-multiplicity subsamples.
And indeed there does appear to be an excess of
low eccentricities in the low-multiplicity systems relative to a smooth
extrapolation of their
higher-eccentricity distributions (see \fig{FracObjEcc}), and this
?excess? of low-eccentricity
orbits is particularly noticeable
in the single-planet and two-planet distributions. This raises the
question of whether or not the turnover in the trend at low-multiplicities 
in
\fig{MeanMedEcc} is real or due to contamination of the low-multiplicity 
by
high-multiplicity systems with so far undiscovered planets.
In the next section we give a crude and model-dependent estimate of
the number of higher-multiple systems contaminating the one- and 
two-planet
distributions by fitting a power law to the higher-multiplicity data.

%%%%%%%%%%%%%%%%%%%%%%%%%%%%%%%%%%%%%%%%%%%%%%%%%%%%%%%%%%%%%%%%%%%%%%
\section{Anaylsis}
\label{sec:Anaylsis}

\begin{figure}[h]
\centering
\includegraphics[scale = 0.68]{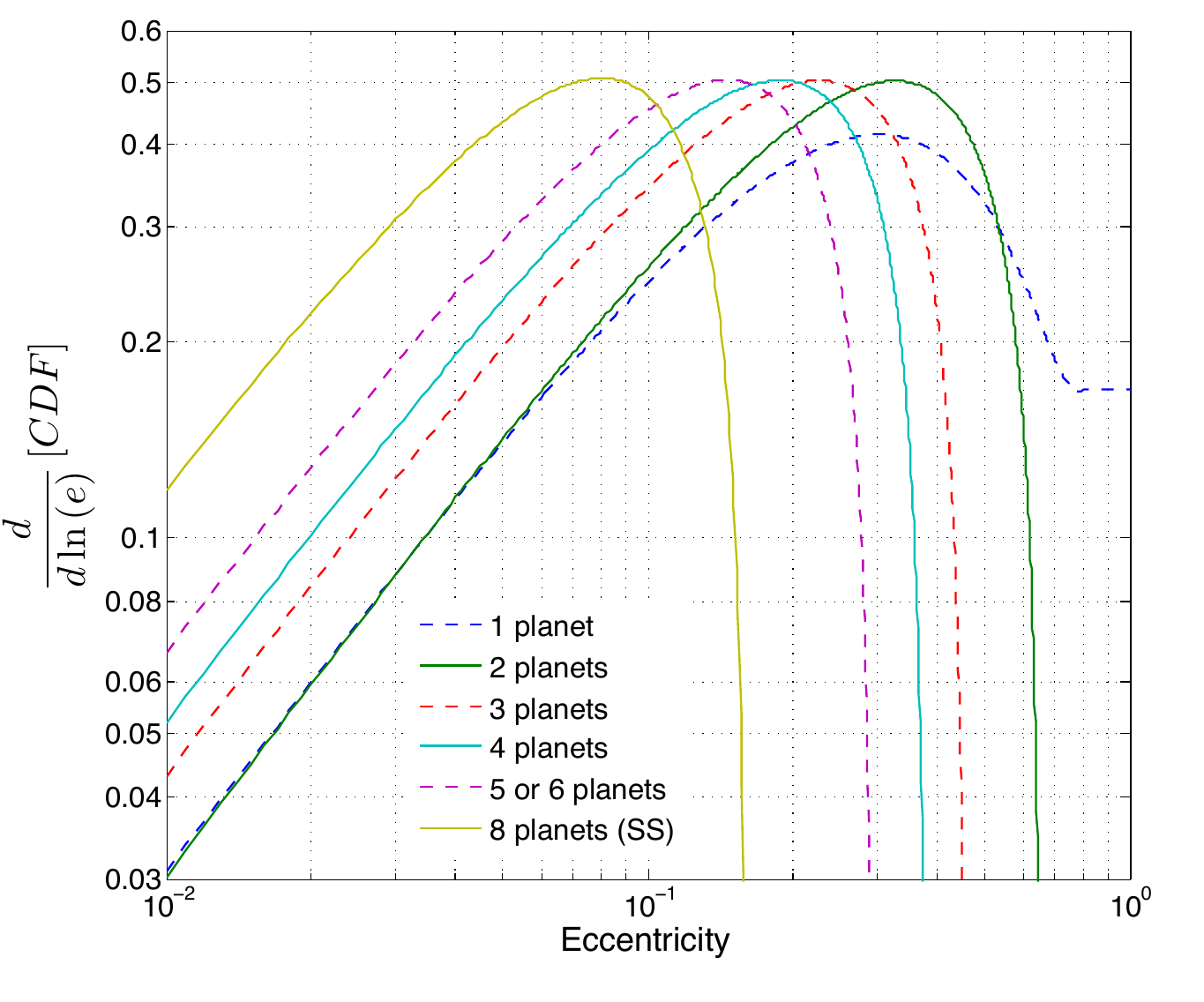}
\caption{Eccentricity probability density distributions for various
multiplicities based on
polynomial fits to the cumulative distribution functions (CDF). Lower
eccentricity systems
are more likely to belong to higher multiplicity systems.  Second-order 
polynomials
were fit to the cumulative distributions for all multiplicities except the
single-planet systems in which case a third-order polynomial was used. The 
`SS'
denotes `Solar System' planets.}
\label{ProbDenLog}
\end{figure}

For each multiplicity, the probability density distributions of
eccentricities were derived
from the cumulative eccentricity distributions and is shown in
\fig{ProbDenLog}. The probability density distributions were obtained by
taking the
derivative of polynomials that were fit to the cumulative eccentricity
distribution
function (the data shown in \fig{FracObjEcc}).  This procedure yields a 
heavily
smoothed estimate
of the true differential distributions.
Second order polynomials were fit to
the data except in the one planet case. A K--S test indicated that a
second
order polynomial was inconsistent (p $<$ 0.01) with the one planet
data, and a
third order polynomial was necessary to obtain an acceptable fit.

If we knew the true occurrence rate of planetary systems as a function of
multiplicity,
these probability density distributions would allow determination of the
likelihood of
a planet with any given eccentricity belonging to a system of a particular
multiplicity.
While that conditional is clearly not satisfied at the present time, it is
clear that if
one wants to find companions to existing exoplanets, one- and two-planet
systems
with low eccentricities are relatively good targets for further
investigation.

\begin{figure}[h]
\centering
\includegraphics[scale = 0.71]{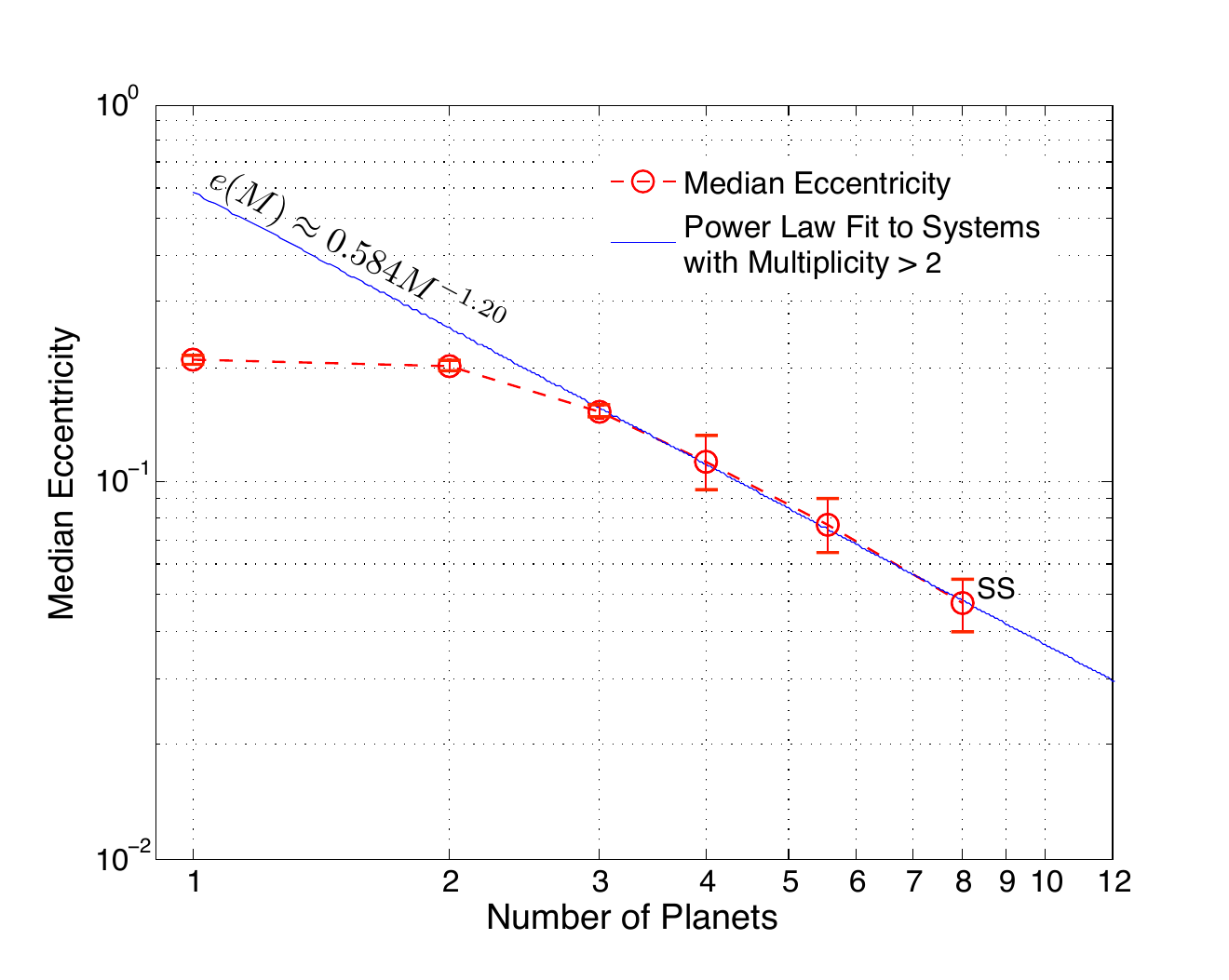}
\caption{Power-law fit to median eccentricity in systems with multiplicity
$>$2
planets. The fit suggests that the one- and two-planet data are 
contaminated
with
higher-multiplicity systems due to so far undiscovered members.  The `SS' 
denotes
`Solar System' planets.}
\label{EccMultFit}
\end{figure}

A power law was fit to the median eccentricity vs. multiplicity
relationship  and is shown in
\fig{EccMultFit}. The fit is only for the higher multiplicity data,
specifically
$M>2$ where $M$ is the multiplicity or number of planets in the system. 
The
one- and
two-planet systems were excluded from the fit on the assumption that they
are likely
contaminated with planets that belong in higher-multiplicity bins
due to so far undetected additional planets in those systems. If the
eccentricity
($e$) verses multiplicity ($M$) relation obeys this power law
\begin{equation} \label{powerlaw}
         e\left(M\right) \approx 0.584M^{-1.20}
\end{equation}
fit, then the relationship can be used to estimate the number of systems
in the one-
and two-planet bins that belong to higher-multiplicity systems. Proceeding
on the
basis of this hypothesis leads to the conclusion that approximately
80\% of the one-planet systems and 25\% of the two-planet systems belong 
to
higher-multiplicity systems with as yet undiscovered members.
While this scenario seems quite plausible, we emphasis that the
estimates of companions in the one- and two-planet bins is dependent on
the assumption that
the true
eccentricity-multiplicity trend behaves like a power law of constant index
at all
multiplicities and that the $M>2$ subsamples are not significantly
contaminated
by yet higher multiplicity systems due to undiscovered members.
It is also quite possible that the turn-over
or plateau seen in the eccentricity distribution is an actual physical
phenomena.

The power law fit can
also be used to predict the median eccentricity of as yet undiscovered
high-multiplicity RV
systems. For example, we estimate that 7-planet RV systems, when
discovered, will
have a median eccentricity of 0.06$\pm$0.01.

In addition to studying the relation between eccentricity and
multiplicity, we also
checked for a trend in semi-major axis with multiplicity but found no
significant
relationship with the exception of the trend noted by 
\cite{Stepinski2000Statistics}
that extremely short period exoplanets ($\lesssim$ 10 days) typically have 
lower
orbital eccentricities, presumably due to tidal circularization.

  \section{Discussion}\label{Discussion}
The distribution of orbital eccentricities as a function of multiplicity
provides an
important new constraint (or clue) for planetary system formation and 
evolution models.
The observational evidence for
the multiplicity-eccentricity relation can perhaps be explained by
planet-planet
interactions and dynamical evolution. The general trend of decreasing
eccentricity
with increasing multiplicity was predicted by \cite{juric2008dynamical} 
using
dynamical evolution simulations. Although the relationship reported here
is qualitatively
similar to that prediction, the observed dependence of $e$ on $M$ is
steeper
and extends to
smaller eccentricities than the one produced by the simulations carried
out by
Juric \& Tremaine \cite{juric2008dynamical}.

Although we are aware of no reason to suspect that the strong dependence
of orbital eccentricity on multiplicity among RV exoplanet systems
reported here is due to some defect or bias in the available data, it is
undeniable that it is based on an inhomogeneous and statistically 
ill-conditioned
data set.  However, even if it were the case that the trend
shown
in \fig{MeanMedEcc} is somehow a spurious non-physical one, the empirical
correlation is still an important one which merits detailed future study
because such a situation would imply that either a large number
of low-eccentricity, low multiplicity exoplanetary systems or 
high-eccentricity,
high-multiplicity ones have been
systematically missed in RV searches.

The rather precise consistency of the eccentricity distribution, as well 
as its mean and median, of the 8-planet Solar System with the observed 
correlations presented above is rather puzzling because existing RV data 
has neither the precision nor the duration required to detect Solar System 
planets other than Jupiter and, perhaps, Saturn.  Thus the ÒfairÓ 
comparison to the RV sample would regard the Solar System as a 
multiplicity one or two case, as illustrated in the figures.  There are at 
least
two possible interpretations:  One is that this is simply a statistical 
fluke and that the comparison of the Solar System to exoplanetary systems 
would be quite different if the RV data used to construct our sample were 
of sufficient quality to
detect all eight Solar System planets.  Another is that the comparison is 
valid because much better RV data would not result in the detection 
of a significant number of additional planets in the $M > 2$ systems.

Intriguingly, the Solar SystemÕs position in eccentricity-multiplicity 
space is not particularly unusual even if it is regarded as an $M = 1$ or 
$M = 2$ case, as illustrated in \fig{eccVSau} which shows that the Solar 
System lies in a reasonably
densely populated region of the space at $M = 1$ or $M = 2$ as well as 
fitting an extrapolation to $M = 8$.

%%%%%%%%%%%%%%%%%%%%%%%%%%%%%%%%%%%%%%%%%%%%%%%%%%%%%%%%%%%%%%%%%%%%%%
\section{Conclusions}\label{sec:Conclusions}
We find that the orbital eccentricities of the Solar System planets are
consistent with those found in
exoplanetary systems when multiplicity is taken into account.
Specifically, we find that as
the multiplicity of a system increases the eccentricity decreases. This
relation can
be well fit by a power law at multiplicities greater than two.
A simplistic and model-dependent interpretation of this fit implies that 
$\sim$80\% of one-planet systems and $\sim$25\% of two-planet systems are 
likely members of
higher-multiplicity systems. The distribution of orbital eccentricities as 
a
function of multiplicity provides an important new constraint for
planetary system formation
and evolution models.  Any theory that accounts for this trend would be
adequate to explain
the distributions of eccentricities seen both in our Solar System and
exoplanetary systems.

Because low eccentricity is arguably advantageous for habitability
\cite{Williams2002Earth-like, Dressing2010Habitable}, this relationship
suggests that
high-multiplicity systems may be more likely to host habitable
exoplanets.

\begin{acknowledgments}
We thank Adam Burrows, AJ Eldorado Riggs and Scott Tremaine for useful
conversations
and comments.  An anonymous refereeÕs comments improved our understanding 
of
valid interpretations of the empirical results presented in the paper.
This research has been supported in part by the World Premier
International Research Center Initiative, MEXT, Japan.
\end{acknowledgments}

\end{article}

   \begin{table}[h]  \label{Table:NumPerBin}
   \begin{center}
    \caption{No. planets in dataset for given multiplicity}
     \begin{tabular}{c c}
  \hline
  \hline
  Multiplicity & Total number of planets  \\
  (no. planets in system):  & with given multiplicity\footnotemark : \\
  \hline
1 & 276 \\
2 & 81 \\
3 & 25 \\
4 & 12 \\
5 or 6 & 9 \\
\hline
\hline
    \end{tabular}
    \end{center}
    \end{table}
\footnotetext{This value is not necessarily a multiple of the system
multiplicity
since not all exoplanets have measured eccentricities.}

   \begin{table}[h]  \label{Table:KStest}
   \begin{center}
   \caption{K--S test on eccentricity for various multiplicities}
   \begin{tabular}{l c c c c c c}
  \hline
  \hline
\multicolumn{7}{c}{P-value for K--S test, including highest 75\% of
eccentricities for each multiplicity
shown in \fig{FracObjEcc}}  \\
  \hline
& 1 planet & 2 planets & 3 planets & 4 planets & 5 or 6 planets & 8
planets \\
1 planet & - &  \\
2 planets & 0.34 & - \\
3 planets & 0.02 & 0.09 & - \\
4 planets & 0.02 & 0.04 & 0.18 & - \\
5 or 6 planets & $<$0.01 & $<$0.01 & 0.05 & 0.46 & - \\
8 planets & $<$0.01 & $<$0.01 & $<$0.01 & 0.05 & 0.09 & -\\
\hline
\hline
\vspace{5mm} \\
  \hline
  \hline
\multicolumn{7}{c}{P-value for K--S test, includes entire sample}  \\
  \hline
& 1 planet & 2 planets & 3 planets & 4 planets & 5 or 6 planets & 8
planets \\
1 planet & - &  \\
2 planets & 0.57 & - \\
3 planets & 0.14 & 0.26 & - \\
4 planets & 0.11 & 0.21 & 0.25 & - \\
5 or 6 planets & 0.02 & 0.02 & 0.16 & 0.64 & - \\
8 planets & $<$0.01 & $<$0.01 & $<$0.01 & 0.12 & 0.12 & -\\
\hline
\hline
    \end{tabular}
    \end{center}
    \end{table}

\begin{thebibliography}{10}

\bibitem{Marcy2005Observed}
Marcy G, Butler RP, Fischer D, Vogt S, Wright JT, Tinney CG, Jones HRA 
(2005)
Observed properties of exoplanets: Masses, orbits, and
  metallicities. { Prog of Theo Phys Sup}~158:24-42.

\bibitem{Udry2007Statistical}
Udry S, Santos NC (2007) Statistical properties of exoplanets. {
  ARA\&A}~45:397-439.

\bibitem{Perryman2011Exoplanet}
Perryman M (2011) {The Exoplanet Handbook.}, Cambridge University Press.

\bibitem{Lissauer1995Urey}
Lissauer JJ (1995) Urey prize lecture: On the diversity of plausible
planetary
  systems. {Icarus}~114:217-236.

\bibitem{Boss1995Proximity}
Boss AP (1995) Proximity of jupiter-like planets to low-mass stars. {
  Science}~267:360-362.

\bibitem{Rasio1996Dynamical}
Rasio FA, Ford EB (1996) Dynamical instabilities and the formation of
  extrasolar planetary systems. {Science}~274:954-956.

\bibitem{Weidenschilling1996Gravitational}
Weidenschilling SJ, Marzari F (1996) Gravitational scattering as a 
possible
  origin for giant planets at small stellar distances. 
{Nature}~284:619-621.

\bibitem{Adams2003Migration}
Adams FC, Laughlin G (2003) Migration and dynamical relaxation in crowded
  systems of giant planets. {Icarus}~163:290-306.

\bibitem{juric2008DynamicalApJ}
Juric M, Tremaine S (2008) Dynamical origin of extrasolar planet 
eccentricity
  distribution. {ApJ}~686:603-620.

\bibitem{juric2008dynamical}
Juric M, Tremaine S (2008) Dynamical relaxation by planet-planet 
interactions
  as the origin of exoplanet eccentricity distribution. {Extreme Solar
  Systems, ASP Conference Series}~398:295-300.

\bibitem{Rodigas2009Which}
Rodigas TJ, Hinz PM (2009) Which radial velocity exoplanets have
  undetected outer companions? {ApJ}~{702}:716-723.

\bibitem{Shen2008OntheEccentricity}
Shen Y, Turner EL (2008) On the eccentricity distribution of exoplanets 
from
  radial velocity surveys. {ApJ}~685:553-559.

\bibitem{Stepinski2000Statistics}
Stepinski T, Black D (2000) Statistics of low-mass companions to stars:
  Implications for their origin. {A\&A}~356:903-912.

\bibitem{Williams2002Earth-like}
Williams DM, Pollard D (2002) Earth-like worlds on eccentric orbits:
  excursions beyond the habitable zone. {Interna J of Astrobio}~1:61-69.

\bibitem{Dressing2010Habitable}
Dressing CD, Spiegel DS, Scharf CA, Menou K, Raymond SN (2010)
  Habitable climates: The influence of eccentricity. {ApJ}~721:1295-1307.

\end{thebibliography}
\end{document}